\documentclass[aps,twocolumn,superscriptaddress,preprintnumbers]{revtex4}
\usepackage{amsfonts,amssymb,amsmath}
\usepackage[]{graphics,graphicx,epsfig}
\usepackage{amsthm}
\usepackage{color}
\usepackage{epsfig}
\usepackage{mathrsfs}
\usepackage{verbatim}
\usepackage{latexsym}

\def\identity{\leavevmode\hbox{\small1\kern-3.8pt\normalsize1}}

\newtheorem{propo}{Proposition}
\newcommand{\be}{\begin{eqnarray} \begin{aligned}}
\newcommand{\ee}{\end{aligned} \end{eqnarray} }
\newcommand{\bpr}{\begin{propo}}
\newcommand{\epr}{\end{propo}}
\newcommand{\bpf}{\begin{proof}}
\newcommand{\epf}{\end{proof}}
\newcommand{\ket}[1]{\left | #1 \right\rangle}
\newcommand{\bra}[1]{\left \langle #1 \right |}

\newcommand{\proj}[1]{\ket{#1}\bra{#1}}

\renewcommand{\epsilon}{\varepsilon}

\def\<{\langle}
\def\>{\rangle}
\usepackage[T1]{fontenc}

\begin{document}

\title{Quantum metrology: Heisenberg limit with bound entanglement}

\author{\L{}. Czekaj}

\affiliation{Institute of Theoretical Physics and Astrophysics, University of Gda\'nsk, 80-952 Gda\'nsk, Poland}
\affiliation{National Quantum Information Centre in Gda\'nsk, 81-824 Sopot, Poland}

\author{A. Przysi\k{e}\.zna}

\affiliation{Institute of Theoretical Physics and Astrophysics, University of Gda\'nsk, 80-952 Gda\'nsk, Poland}

\author{M. Horodecki}

\affiliation{Institute of Theoretical Physics and Astrophysics, University of Gda\'nsk, 80-952 Gda\'nsk, Poland}
\affiliation{National Quantum Information Centre in Gda\'nsk, 81-824 Sopot, Poland}

\author{P. Horodecki}

\affiliation{Faculty of Applied Physics and Mathematics, Gda\'nsk University of Technology, 80-233 Gda\'nsk, Poland}
\affiliation{National Quantum Information Centre in Gda\'nsk, 81-824 Sopot, Poland}

\begin{abstract}
Quantum metrology allows for a huge boost in the precision of parameters estimation. However,  it seems to be extremely sensitive on the noise.
Bound entangled states  are states with large amount of noise what makes them unusable for almost all quantum informational tasks.
Here we provide a counterintuitive example of a family of bound entangled states which
can be used in quantum enhanced metrology.
We show that these states give advantage as big as maximally entangled states and asymptotically reach the Heisenberg limit.
Moreover, entanglement of the applied states is very weak which is
reflected by its so called unlockability poperty.
Finally, we find instances where behaviour of Quantum Fisher
Information reports presence of bound entanglement while a well-known class of strong correlation Bell inequality
does not.  The question rises  of  whether (and if so, then to what degree)
violation of local  realism is required for the sub-shot noise precision in quantum metrology.
\end{abstract}

\maketitle

\emph{Introduction.}
Estimation of a physical parameter is an important goal in many areas of science \cite{Advances}.
One of intriguing aspects of quantum mechanics in this context is quantum metrology.
In its most popular form quantum metrology has its origins in atomic spectroscopy  \cite{Bollinger,Huelga},
however, the idea was present even earlier on the ground of fermionic systems \cite{Yurke} and
- from different perspective - quantum optical interferometry \cite{Caves} (for
recent development along this line see \cite{Benatti} and references therein).
General search for possible improvement of estimation precision is also of great importance
in the domain of atomic clocks \cite{Buzek} and in interferometric measurement of a phase shift in gravitational wave detection \cite{LIGO}.
Since there are many quantities we cannot measure directly, a protocol of a measurement is typically indirect - we use an additional probe system to interact with the one under investigation.
Due to the interaction the probe gains information about the parameter we want to measure.
Then we inspect the probe coming out from the measurement and, basing on obtained data, we estimate the desired parameter.
Obviously, we want to obtain the highest possible accuracy of that estimation.
We can improve the accuracy repeating the experiment multiple times or, equivalently, make multipartite probe to interact with the system.

In the quantum world there is another possibility of increasing the accuracy: prepare the probe in a particular quantum state i.e. in the entangled state.
To be more concrete, for a classical probe that contains $n$ particles (we can also consider it as a measurement performed $n$ times) accuracy scales like  $ 1/\sqrt n$. That is so called Shot-Noise Limit (SNL).
However, if the system is in particular entangled state, then accuracy can be improved up to $1/n$. This limit, called Heisenberg Limit (HL) gives us the best what we can get that is allowed by quantum mechanics.
Both of these bounds can be derived from quantum Cramer-Rao bound and Quantum Fisher information (QFI) \cite{QFI_review,QFI1,QFI2,QFI3,QFI4}.
Recently it has been shown \cite{DD} that local (memoryless) noise puts the limit on the accuracy
offering only  linear improvement of precision if compared to the shot-noise limit.
The above, however, does not nullifies all possible advantages of the phenomenon.
First, there is an extra option to filter out some part of noise with help of dedicated
error correction schemes (see \cite{QErrorMetrology1,QErrorMetrology2,QErrorMetrology3}).
Second,
any measurement resulting in improvement below shot-noise limit
is a signature of quantum entanglement in the system, that is quantum metrology technique
may serve as a witness of genuine quantum correlations in the system.
Indeed, how powerful it may we, we shall see below.

It is known that genuine multipartite quantum entanglement is necessary  to surpass the SNL (see \cite{Toth}),
however not every entangled state gives the same improvement, and among entangled states there are also states that are not suitable for quantum metrology i.e. they do not surpass SNL. In particular quantum scaling is hard to obtain in case of entangled states with high noise factor (see \cite {DD}) which we have to deal with in realistic experiments where decoherence and preparation errors are present.

States on which we focus in this paper belong to  a group of states with such high noise factor that makes them unusable for most of the quantum information tasks. These highly mixed states are bound entangled (BE) states and were predicted in 1998 \cite{Distillation, BE} as a new kind of entanglement.
Bound entangled states are those from which no pure entanglement can be distilled when only local operations and classical communication (LOCC) are available.
The sufficient condition for entangled state to be bound entangled is its positive partial transposition \cite{PPT1,PPT2}.
The bound entangled states, called "black holes" of quantum information \cite{Terhal1}, have been created in laboratories in a series of experiments with ions, photons and nuclear spins.
Among multipartite bound entangled  states we distinguish unlockable and non-unlockable ones.
Unlockable BE states are those, in which grouping some parties together and allowing them to  perform collective quantum operations, makes distillation of pure entanglement between two other parties outside the group possible.
Non-unlockable BE states are those in which we cannot obtain pure entangled state by these means. One may
say that entanglement is "more bound" there.

Impossibility of  pure entanglement distillation makes BE states (in particular non-unlockable ones) not useful for many quantum information and communication tasks
such as quantum teleportation or dense coding. When it comes to quantum cryptography,
on the one hand, it has been shown that BE states may be useful \cite{BoundCrypto}, on the other hand very recent results show that the resulting cryptographic key in some cases may be not suitable for quantum repeater schemes \cite{CryptographyRepeaters}.
In case of metrology, no instance of usefulness of BE states was known so far. In \cite{Laskowski} the authors relate QFI and BE states and show that for certain BE states, averaged Fisher Information is higher than for separable states.
However, even though the relation with averaged QFI was given, the usability of BE in case of standard formulation of quantum metrology (i.e. with known interaction between a system and a probe) remained an open question.

Our motivation here is
the lack of knowladge which states are useful for the
standard quantum metrology and which are not.
Fitness of BE for purposes of quantum information theory (especially in the context
of quantum information processing) is also not fully recognized yet.
For these reasons we focus on the long-standing question "Do, among bound entangled states, there exist any examples  that beat the shot-noise limit?".
Intuition suggests that the high degree of noise of BE should be the reason of the negative answer.
It is, in particular,  especially tempting to expect such answer
in classes of multiqubit states, the entanglement of which can not be unlocked.

Indeed, the well known fact that any qubit-qudit system violating positive partial transpose condition
is free entangled has a immediate consequence for lockability of n-qubit systems: unlockable
multiqubit bound entanglement must {\it satisfy positivity partial transpose condition with respect to any single qubit}. On the other hand, it is known that partial
transpose has some relation to local time reversal.
Since quantum metrology is about estimation of
parameter with respect to speed of the evolution
in time, it might be tempting to expect that this
type of entanglement should be very hard or even
impossible to exhibit metrology below shot-noise
limit.

This intuition is, however,  misleading. We investigate a class of mixed states that are GHZ-diagonal and present the first, to our knowledge, example of bound entangled states which have advantage over product states in metrology of phase shift around z-axis. What is more, in the discussed states, the entanglement cannot be unlocked.
Our family of states exhibit an $a n^{2}$ scaling  (with $a\ge \frac{1}{4}$) 
of the QFI in the asymptotic limit.
We compare QFI with multipartite Bell inequalities (as a tool of entanglement detection) and find that in some cases the sub-shot noise reports entanglement even when the well-known rich class of correlation Bell inequalities do not.

\emph{Quantum Fisher Information for GHZ-diagonal states.}
We consider a class of $n$-qbit states that are diagonal in the generalized GHZ-basis:
\begin{equation}
\rho=\sum_{i=0}^{2^{n-1}-1}{\left( \lambda_i^+\proj{\phi^+_i}+\lambda_i^-\proj{\phi^-_i}\right)}\label{rhoGHZ},
\end{equation}
where for simplicity we assume that $\lambda_i^+ \ge \lambda_i^-$.
States $\rho$ constitute superset of states studied in \cite{DCprl, DCactivating, DCclassification} in the context of separability and distillability conditions.
By generalized GHZ-basis we mean:
\begin{equation}
\ket{\phi^\pm_i}=\frac{1}{\sqrt{2}}\left(\ket{i}\pm\ket{\bar{i}}\right),
\label{phi}
\end{equation}
where for $n$-qubit system  $i \in \{0,1,..., 2^{n-1}-1\}$.
Here we put $n$-digit binary representation of $i$ in $\ket{i}$ and its negation in  $\ket{\bar{i}}$.
Note that in the range of indices (i.e. $\{0,1,..., 2^{n-1}-1\}$) the $n$-digit binary representation of $i$ always starts with $0$.
For example, for $4$-qubit system we have $\ket{\phi^\pm_2} =\frac{1}{\sqrt{2}} \left(\ket{0010}\pm\ket{1101}\right)$.

We study usefulness of states $\rho$ for quantum metrology in terms of Fisher Information (FI).
FI quantifies the amount of information on unknown parameter $\theta$ that may be extracted by measurements.
For a probe state $\rho(\theta)$ which depend on the parameter $\theta$ and the positive-operator valued  measurement (POVM) with elements $\{E_\mu\}$ and values $\mu$, FI reads:
\begin{equation}
F=\sum_\mu \frac{1}{P(\mu|\theta)}[\partial_\theta P(\mu|\theta)]^2,
\end{equation}
where $P(\mu|\theta)=Tr[\rho(\theta) E_\mu]$ are conditional probabilities and
POVM values $\mu$ estimate the parameter $\theta$. FI gives a lower bound for a standard deviation of the estimator for fixed value of the parameter $\theta$ \cite{cramer_rao}:
\begin{equation}
\Delta \theta_{est}=\sqrt{ \< (\mu-\theta)^2 \> }\ge\frac{1}{\sqrt{F}}.
\label{Delta}
\end{equation}
The maximum value of FI which may be achieved by measurement optimisation \cite{QFI1,QFI2,QFI3,QFI4} is given by the quantity called Quantum Fisher Information (QFI). It depends only on the initial state of the probe system and a form of evolution which links estimated parameter $\theta$ and the final state of the probe system $\theta\mapsto\rho(\theta)$.
In  case of multipartite separable states, maximal value of QFI scales linearly with the system size (SNL).
This is reflected by the separability condition for quantum Fisher information $F_{Q}$ (see \cite{Laskowski}), i.e. for any separable state $\rho_{sep}$, it holds:
\begin{equation}
F_{Q}(\rho_{sep}) \leq n.
\label{QF-separability-test}
\end{equation}
On the contrary, the highest scaling i.e. quadratic one (HL) $F_{Q}(\rho) \approx n^{2}$ may be achieved only by entangled states $\rho$.
For more introduction see \cite{QFI_review}.

In this paper we discuss a setup where a phase shift around z-axis is estimated. Probe state $\rho$ undergo evolution according to $U_\theta=\exp\left[-i \theta Z\right]$ where $Z$ is a Hermitian generator of the form $Z=\left(\sigma_Z^{(1)}+\ldots+\sigma_Z^{(n)}\right)/2$. Index $^{(i)}$ denotes a qubit on which $\sigma_Z$ acts.
In such case, QFI for the probe state $\omega=\sum_i \lambda_i \proj{\phi_i}$ is given by \cite{Braustein}:
\begin{equation}
F_Q= 2 \sum_{i,j} \frac{(\lambda_i -\lambda_j)^2}{\lambda_i + \lambda_j}\left|\bra{\phi_i}Z\ket{\phi_j}\right|^2.
\label{FQgen}
\end{equation}
For the states of the form \eqref{rhoGHZ}, this formula simplifies to:
\begin{equation}
F_Q= \sum_{i}w_i^2 \frac{(\lambda_i^+ -\lambda_i^-)^2}{\lambda_i ^++ \lambda_i^-},
\label{FQdiag}
\end{equation}
where $w_i=(\#0(i)-\#1(i))$ is the difference between number of "0"s and "1"s in binary representation of $i$ (e.g.  in $4$-qubit system $w_{0}=4$,$w_{1}=2$,$w_{2}=2$ etc.).
This is easy to check when we observe that operator $Z$ is diagonal in standard basis with $z_{i,i}=w_i, z_{2^n-1-i, 2^n-1-i}=-w_i$.
and that only nonzero terms are those with $\bra{\phi_i^+}Z\ket{\phi_i^-}=w_i/2$.

\emph{Bound entangled GHZ-diagonal states.}
Since here we are interested in bound entangled states, we derive criterion for a state to be BE.

\begin{figure}
\includegraphics[width=0.45\textwidth]{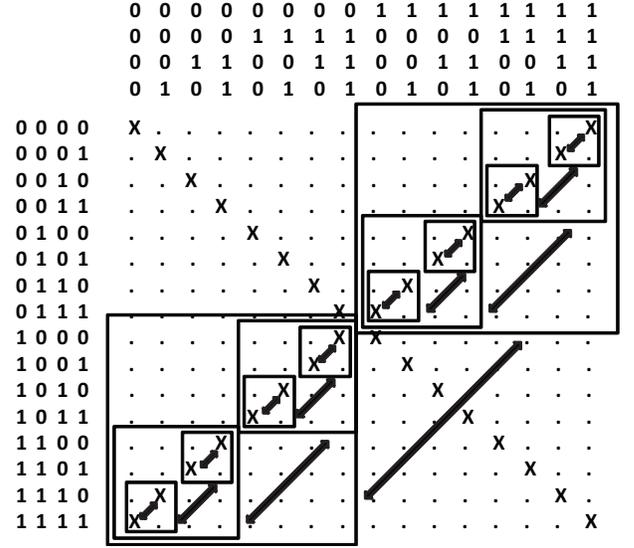}
\caption{The pictorial structure of exemplary (four qubit) GHZ diagonal state written
in the standard basis (the numeration of the basis is explicitly provided).
Only the elements depicted by ''$X$'' are nonzero.
Partial transposition
with respect to a given qubit is represented by the group of
elementary transpositions represented by  arrows of the same length
starting from the longest arrow that corresponds to the transposition
with the first qubit. Note that partial transpositions with respect to different
qubits commute. Hence, a transposition with respect to a given subset of qubits corresponds
just to a composition of the several operations, each consisting of all arrows of one kind.
For example the state transposed partially with respect to the first and the last
qubit results form the application of the operation corresponding to
the longest arrow followed by the operations corresponding to all the smallest arrows.
}
\label{Transposition}
\end{figure}

\begin{propo}
A GHZ-diagonal state is unlockable bound entangled for every $ 1:(n-1)$  cut  if its eigenvalues $\lambda^\pm_i$ satisfy
\begin{equation}
\min_{i\in\Omega_j}( \lambda_i^+ + \lambda_i^-)  \ge \left| \lambda_j^+ -\lambda_j^-\right|,
\label{PPTcond}
\end{equation}
for every $j\in{0,1,...,2^{n-1}-1}$, where
\begin{equation}
 \Omega_j=\{NOT_{2,3,...,n}(j)\}\cup
\{ NOT_k(j)| k\in\{1,2,...,n\}\}.
\end{equation}
$NOT_k(j)$ is a negation on the $k$-th bit of the binary representation of $j$.
\end{propo}
Before proving the above let us remind that the GHZ diagonal family has  a diagonal-antidiagonal
form and the partial transpose with respect to each qubit looks eminently simple here (see Fig. \ref{Transposition}
for illustration).
Let us now proof the above Proposition.
\bpf
First we show that the state is PPT for every $1:(n-1)$  cut.
GHZ-diagonal state written in the standard basis contains nonzero elements only on diagonal and antidiagonal: $\rho_{i,i}=\rho_{2^n-1-i,2^n-1-i}=\frac{\lambda_i^++\lambda_i^-}{2}$ and $\rho_{i,2^n-1-i}=\rho_{2^n-1-i,i}=\frac{\lambda_i^+-\lambda_i^-}{2}$.
Partial transposition with respect to $k^{th}$-qubit influences only antidiagonal elements (in the standard basis) of the density matrix such that
\begin{equation}
\lambda_i^+-\lambda_i^- \rightarrow \lambda_j^+-\lambda_j^-,
\end{equation}
where
\begin{equation}
i=\left\{\begin{array}{ll}NOT_{2,3,...,n}(j) &\text{ if\ \ \ } k=1, \\
NOT_k(j) &\text{ elsewhere.}\end{array}\right. \label{transp}
\end{equation}
 The state after the transposition remains diagonal-antidiagonal and its eigenvalues are
\begin{equation}
\Lambda_i^{k \pm}=\frac{1}{2}\left[(\lambda_i^++\lambda_i^-)\pm (\lambda_j^+-\lambda_j^-)\right].
\end{equation}
They are positive when:
\begin{eqnarray}
\lambda_i^++\lambda_i^-\ge  \left| \lambda_j^+ -\lambda_j^-\right|.
\label{ai}
\end{eqnarray}
Finally state is PPT with respect to every 1-qubit partial transpositions when:
\begin{equation}
\min_{i\in\Omega_j} ( \lambda_i^+ +\lambda_i^-)\ge \left| \lambda_j^+ -\lambda_j^-\right|,
\label{BEcond}
\end{equation}
for $\Omega_j=\{NOT_{2,3,...,n}(j)\}\cup
\{ NOT_k(j)| k\in\{1,2,...,n\}\}$ what comes from condition \eqref{transp}.
If the state is PPT for given $ 1:(n-1)$  cut, no entanglement can be distilled between these two parties.
Since we put $n-1$ parties together in this cut, every collective quantum operation is allowed for this group.
It is easy to see that this setup is less restrictive than unlocking protocol.
That means that two particle entanglement can not be unlocked for any $2$ particles of the discussed states.
\epf
\emph{The family of $\rho_{n,k}$ states.}
Here we introduce a subset of GHZ-diagonal states. It contains states $\rho_{n,k}$ that are BE and, as we show next, are useful for quantum metrology.
The considered states $\rho_{n,k}$ are parameterized by two numbers: $n$ which is the number of qubits in the system, and $k$ that
characterises the structure of the states as follows:
\begin{equation}
\rho_{n,k}=\lambda P_{n,k}^{+} + \frac{\lambda}{2} (Q_{n,k}^{+} + Q_{n,k}^{-})
\end{equation}
where the projector $P_{n,k}^{+}=\sum_{i: \ \#1(i)<k \text{\ \ or\ \ } \#1(i)> n-k} \proj{\phi^+_i}$ and
 $Q_{n,k}^{\pm}=\sum_{i: \ \#1(i)=k \text{\ \ or\ \ } \#1(i)= n-k} \proj{\phi^\pm_i}$
with the linear factor $\lambda=1/\sum_{i=0}^{k}\binom{n}{i}$ following directly
from the normalisation condition $Tr(\rho_{n,k})=1$. Here the notation $\#1(i)$ means just the
number of ones in the binary representation of the number $i$. For instance $\#1(i=7)=3$ since
the binary representation "1101"  of the number $7$ contains three ones.
The exemplary state $\rho_{n,k}$ with $n=4,k=2$ is presented below
\begin{equation}
\rho_{4,2}=\frac{1}{11}\sum_{i\in{\it I}_{1}}{\proj{\phi^+_i}}+ \frac{1}{22} \sum_{i\in{\it I}_{2} }{\proj{\phi^+_i}+ \proj{\phi^-_i}}
\label{rho42}\end{equation}
where ${\it I}_{1}=\{0,1,2,7\}$ and  ${\it I}_{2}=\{ 3,4,5,6 \}$.
\begin{propo}
For any  $n,k$ the state $\rho_{n,k}$ passes positive partial transpose test (PPT) with respect to local transposition
on any single qubit system and, as such, it is bound entangled.
\end{propo}

\bpf
To proof that $\rho_{n,k}$  is BE we have to show that it satisfies \eqref{PPTcond} for every  $n$ (number of qubits) and $k$ (maximal number of ones in binary representation of indices $i$ associated to nonzero eigenvalues).
From conditions \eqref{transp} we can see that, for a given $i$,  $\Omega_j$ contains numbers which binary representation  have only three possible numbers of ones: $j+1$,  $j-1$, $n-j-1$.
Let us discuss separately three different possibilities:
a) $ \#1(j)<k $ or $ \#1(j)>n-k$. In $\Omega_j$ we can have only those $i$ with  $\#1(i)\le k $ or $\#1(i)\ge n-k$. Therefore $\min_{i\in\Omega_j} ( \lambda_i^+ +\lambda_i^-)=\lambda$ and $ \left| \lambda_j^+ -\lambda_j^-\right|=\lambda$ and \eqref{PPTcond} is satisfied. b) $\#1(j)=k $ or $ \#1(j)=n-k$.
Then in $\Omega_k$ we  have  $i$ with  $\#1(i)\in\{ k-1, k+1, n-k-1\}$.
In $\Omega_{n-k}$ there are $i$ satisfying $\#1(i)\in\{ k-1, n-k-1, n-k+1\}$.
In both cases  $\min_{i\in\Omega_j} ( \lambda_i^+ +\lambda_i^-)=\lambda$ and $ \left| \lambda_j^+ -\lambda_j^-\right|=0$ and \eqref{PPTcond} is satisfied.  c)  $ \#1(j)>k $ or $ \#1(j)<n-k$. In $\Omega_j$ we can have only those $i$ with  $\#1(i)\ge k $ or $\#1(i)\le n-k$.
 With at least one $i$ for which these inequalities are sharp. So $\min_{i\in\Omega_j} ( \lambda_i^+ +\lambda_i^-)=0$ and $ \left| \lambda_j^+ -\lambda_j^-\right|=0$ which also satisfy  \eqref{PPTcond}. With the above we have checked all the transposition for any given $n,k$ and seen that indeed the \eqref{PPTcond} is always satisfied which proofs that $\rho_{n,k}$ states are non-unlockable  BE.
\epf

It is easy to check that state $\rho_{n,k}$ is NPPT in the $m:(n-m)$ cuts  for $m\geq 2$.

\begin{figure}
\includegraphics[width=0.5\textwidth]{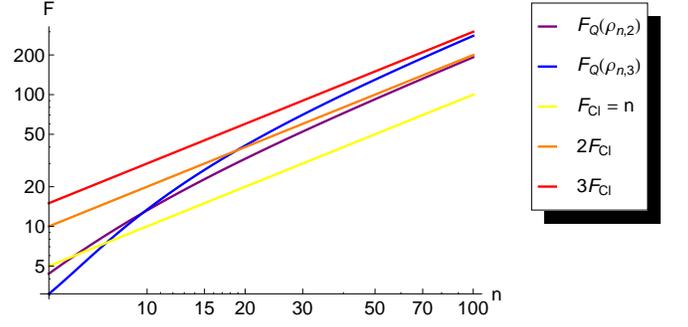}
\caption{Precisely calculated QFI and the limit it achieves in the infinity for two different values of $k$. Blue  and purple lines show how the $F_Q^{n,k}$ change for $k=3$ and $k=2$ respectively and red and orange lines are limits that these functions achieve in the infinity. }
\label{k23}
\end{figure}

{\it Scaling of Quantum Fisher Information.}
In case of the states under consideration, with the assumption $k<\lfloor \frac{n}{2}\rfloor$,  the equation for Quantum Fisher Information \eqref{FQdiag} takes the form:
\begin{equation}
F_Q^{n,k}= \lambda \sum_{j=0}^{k-1}(n-2j)^2\binom{n}{j}.
\label{FQrho}
\end{equation}
It simply comes from counting number of states with given $w$.
In picture \ref{k23} we show how QFI approaches limit of $nk$ for $k=2$ and $k=3$.
We see that if the $k$ does not grow with $n$ the usual shot-noise classical limit is kept.
However the things dramatically change when we put dependence of $k$ on $n$.
In what follows, we shall utilise this fact proving the central result of the paper
\begin{propo}
Quantum Fisher Information $F_Q^{n,k}$ given by equation (\ref{FQrho}) satisfies
\begin{equation}
F_Q^{n,k}\geq (n-2k)^{2}\frac{k}{n+1}
\label{lower-bound}
\end{equation}
for any n and $k<\frac{n}{2}$. In particular putting $k(n)=a n$ ($a < \frac{1}{2}$)
it follows the asymptotic behaviour
\begin{equation}
\lim_{n \rightarrow \infty} \frac{F_Q^{n,k(n)}}{a(1-2a)n^{2}}\geq 1.
\label{limit}
\end{equation}
\end{propo}
The limit (\ref{limit}) immediately result from (\ref{lower-bound}).
Here we only  derive (\ref{lower-bound}) for the quantum Fisher information in the form (\ref{FQrho}):
\bpf
First, we bound the QFI from below:
\begin{eqnarray}
F^{n,k}_{Q}&=& \frac{ \sum_{j=0}^{k-1}(n-2j)^2\binom{n}{j}}{ \sum_{j=0}^{k}\binom{n}{j}} \nonumber \\
&\ge&(n-2k)^2\frac{ \sum_{j=0}^{k-1}\binom{n}{j}}{ \sum_{j=0}^{k}\binom{n}{j}}
\label{Flim1}
\end{eqnarray}
Consider the last factor $S_{n,k}:= \frac{\sum_{j=0}^{k-1}\binom{n}{j}}{ \sum_{j=0}^{k}\binom{n}{j}}$.
One can estimate its inverse $S_{n,k}^{-1}$ as follows:
\begin{eqnarray}
&&S_{n,k}^{-1}=\frac{\sum_{j=0}^{k}\binom{n}{j}}{\sum_{j=0}^{k-1}\binom{n}{j}}= \frac{ \sum_{j=0}^{k-1}\binom{n}{j} + \binom{n}{k}}{ \sum_{j=0}^{k-1}\binom{n}{j}} = \nonumber \\
&& = 1 + \frac{\binom{n}{k}}{ \sum_{j=0}^{k-1}\binom{n}{j}} \leq 1 + \frac{\binom{n}{k}}{\binom{n}{k-1}}=\frac{n+1}{k}
\end{eqnarray}
and hence the original factor satisfies $S_{n,k} \geq \frac{k}{n+1}$ from which the lower bound (\ref{lower-bound})
immediately follows.
\epf

\begin{figure}
\includegraphics[width=0.5\textwidth]{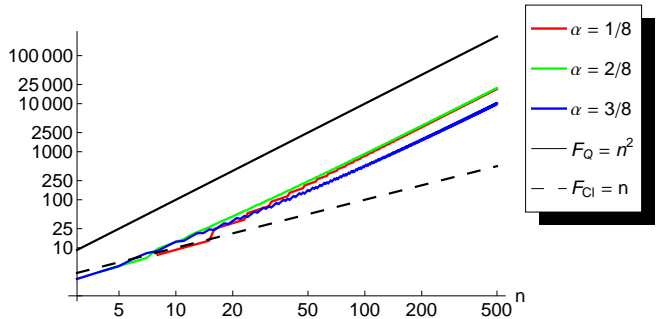}
\caption{Asymptotic behaviour of the quantum Fisher information for $\rho_{n,k}$ states in the case when $k=a n$. Dependence on $n$ is calculated for three different
values of $a$: 1/8 (blue dots), 1/4 (green dots) and  3/8 (red dots).}
\label{alpha_large_scale}
\end{figure}

{\it Remark (exact form of the Heisenberg limit) .} Optimising the denominator
in the formula (\ref{limit}) gives
 $a=\frac{1}{4}$. Hence, the Fisher information
scales in this case not worse than $\sim \frac{n^{2}}{8}$. In the end, this  results in an upper bound for the scaling of the measurement precision (\ref{Delta})
by $\frac{2\sqrt{2}}{n}$ as apposed to its shot noise lower bound $\frac{1}{\sqrt{n}}$.
Asymptotic behaviour of $F_Q^{n,k(n)}$ is illustrated in the picture \ref{alpha_large_scale} for
three different values of $a$.

\emph{The relationship to the Bell inequalities: sub-shot noise
precision does not require violation of local realism?}
One of the fundamental features of
quantum states caused by quantum entanglement is {\it lack of local realism}.
It is known that some entangled states satisfy
all Bell inequalities since the explicit hidden variable
model can be constructed for them (see \cite{Werner}
for some states with nonpositive partial transpose
and \cite{Laskowski} for PPT states).
This, however, does not follow automatically that
the states  are fully locally realistic.
As it was shown recently
\cite{Brunner_2},
there exist entangled states, the nonlocality of which can be revealed only by using a {\it sequence of measurements} (i.e. when each party performs sequentially more than one measurement on the system, for formal definitions see \cite{sequential1,sequential2}, it was formerly discussed in Refs. \cite{Popescu,Gisin}.
One of the fundamental open question of quantum physics is
 whether all entangled quantum states are nonlocal at least in the latter, weaker, sense.

\begin{figure}
\includegraphics[width=0.5\textwidth]{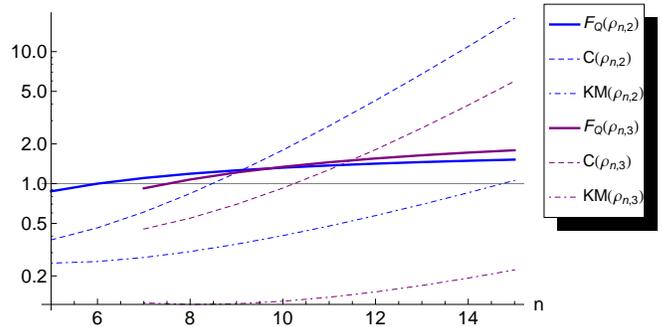}
\caption{Comparison of Fisher Information criterion and correlation condition in the
power of entanglement detection. Here we plotted $F_Q/F_{Cl}$ and
$\mathcal{C}^{(n)}$ for two classes of states $\rho_{n,2}$ and
$\rho_{n,3}$. Tests detect entanglement when their values exceed $1$.
The most interesting region is where Fisher Information criterion detect
entanglement for states with hidden variable model for two
dichotomic observables on each site
(i.e. $F_Q/F_{Cl}>1$ and $\mathcal{C}^{(n)}<1$).
For comparison we also plot value of Klyachko-Mermin (KM) inequality. For
analysed states it performs much worse than $\mathcal{C}^{(n)}$.}
\label{entanglement-detection}
\end{figure}

Here we find that the the efficency of Fisher information separability test
does not coincide with the efficiency of some strong Bell inequality tests.
We have chosen the family of all $n$-qubit-correlation Bell inequalities
with   $2^{n-1} \times 2^{n-1} \times ... \times 2^{n-2} \times ... \times 2$
settings per sides (see \cite{BruknerZL}, \cite{PaterekLZ}), which  can be written as
the simple inequality
\begin{equation}
{\cal C}^{(n)}(\rho) \leq 1
\label{Bell}
\end{equation}
where ${\cal C}^{(n)}$ is a special (optimised) function of
the correlation tensor of the state (see Appendix).
The above can be seen as a necessary condition for separability of
any $n$-qubit state.
Remember that another necessary condition for separability of
$n$ qubits is the shot-noise limit bound (\ref{QF-separability-test}).
We have calculated the upper bound (see Appendix) for the factor
(\ref{Bell}) for some states form the class $\rho_{n,k}$
or some states from the
class $\rho_{n,k}$. We obtained that for $k=2$ and $k=3$ and $n$
respectively from the set $\{7,8\}$ and $\{8,9,10\}$, Fisher Information
criterion outperform correlation Bell inequality condition (\ref{Bell}),
i.e. it detects entanglement while correlation condition indicate hidden variable model (see the
Figure (\ref{entanglement-detection})for the details).
This is surprising since quantum metrology involves only two setting of binary Pauli observables per site.
We believe that the possible power of QFI lies in {\it its differential character}. Indeed, the above observation suggests
need of deep study of nonlocality in the context of metrology which was, to our knowledge, not pursued so far.
In particular one of the questions that may be raised is possible role of metrology as a necessary condition for standard, or even weaker i.e. sequential,
nonlocality. In fact it may happen that some of the presented states even allow for the general (not only finite number of settings, like here, but with continuum settings)
single measurement hidden variable model, nevertheless exhibit sequential nonlocality.
Metrology may be just a first signature of that.

\emph{Discussion and conclusions.-}

We have shown by explicit construction that there exist bound entangled states that can be useful
for quantum metrology  and can reach the accuracy scaling  exactly according to the Heisenberg limit.
The result has been shown for a very noisy entanglement which is not only unlockable but even essentially
weaker since as long as some qubit is kept as an elementary subsystem
no collection of the remained qubits into groups can result in free entanglement.

The most natural question here is about the maximal value of the linear factor
in the sub-shot noise limit for bound entanglement $c n^{2}$. We have found
here that $c$ is not less than $\frac{1}{8}$. The question is whether
it can reach the optimal pure GHZ state value and only the speed of the convergence
is an issue, or there is a threshold imposed on $c$ from bound entanglement property.

The character of the results opens new directions of possible research.
The fact that Fisher information outperforms strong Bell inequalities as a multiparty
entanglement witness naturally suggests need of further analysis of the interplay of
the role of nonlocality in the sub-shot noise limit.
The natural question (especially in the context of recent results on nonlocality \cite{Brunner_2}) to answer
would be: is there any family of quantum states that allows for
general LHV model, but can be used to obtain sub-shot noise (ie. better than classical) quantum metrology?
It is related to another one (especially in context of
both general requirements in quantum metrology \cite{Toth} as well as recent results on nonlocality \cite{Brunner_2})
whether there is any chance for sub-shot noise metrology for states obeying the PPT condition with respect to {\it any} cut?
While the present result may be generalised to get the sub-shot noise metrology 
with bound entanglement having PPT property under arbitrary sublinear fraction of qubits,
further improvement to PPT under any cut seems to be not possible for GHZ-diagonal states.
We believe, however, that our result can be generalised to Dicke-type states, where possible 
chance for positive answer to the above question may be more likely.


Next  important question that naturally arises - especially because of the
unlockability property - is : What is the general role of error correction in case of metrology?
In fact the noise can never be filtered out form bound entanglement
and this is the reason why the corresponding binding entanglement channels
are resistant to {\it any} error correction and have quantum capacity zero \cite{binding1}, \cite{binding2}.
This is what makes the present result quite nonintuitive.


As a by-product, it seems that the phenomenon presented here may reopen the fundamental question
of quantum computational tasks at a high noise rate even on the level of quantum correlations
beyond entanglement \cite{JozsaLinden}. Indeed, as apposed to the
quantum games theory based on ,,kinematical'' aspects of quantum physics, quantum metrology exploits
dynamics explicitly and, as such, may be closer to the perspective of quantum algorithmic tasks.


\emph{Acknowledgements.}
We thank Pawe\l{} Mazurek for numerous discussions
at early stage of this project.
This work is supported by ERC AdG QOLAPS and EC grant QUASAR.
AP is supported by the International PhD Project ``Physics of future quantum-based information technologies''
grant MPD/2009-3/4 from Foundation for Polish Science and by the University of Gda\'nsk grant 538-5400-B169-13.

\section{Appendix}

{\it Correlation Bell inequalities .} The necessary and sufficient condition for two-settings correlation Bell inequalities
\cite{BruknerZukowski,WernerWolf} (including the Mermin-Klyshko inequality \cite{MKBell})
has been later extended then to {\it necessary and sufficient condition}
for $2^{n-1} \times 2^{n-1} \times ... \times 2^{n-2} \times ... \times 2$
settings correlation Bell inequalities \cite{BruknerZL}, \cite{PaterekLZ}. The condition,
is given by a specific formula for a correlation tensor in the form:
\begin{equation}
{\cal C}^{(n)}(\rho)= max \sum_{k_{1}, ...k_{n}=1}^{2} T_{k_{1}...k_{n}}^{2} \leq 1
\label{inequality}
\end{equation}
where $T_{k_{1}...k_{n}}=Tr[\hat{e}_{k_1} \hat{\sigma} \otimes  ... \otimes \hat{e}_{k_n} \hat{\sigma} \rho]$
and the maximum is taken over all possible sequences of pairs of orthogonal vectors $\{ \hat{e}_{k_i}, k_{i}=1,2\}_{i=1}^{n}$.
Quite remarkably the LHS of the above satisfies the condition
\begin{equation}
{\cal C}^{(n)}\leq ||T||_{HS}^{2}=\sum_{k_{1}, ...k_{n}=1}^{3} T_{k_{1}...k_{n}}^{2}
\label{bound}
\end{equation}
since the latter is a square of the Hilbert-Schmidt norm of the operator
$T=\sum_{k_{1}, ...k_{n}=1}^{3} T_{k_{1}...k_{n}}^{2} \hat{e}_{k_1} \hat{\sigma} \otimes  ... \otimes \hat{e}_{k_n} \hat{\sigma}$
(representing the correlation part of the state $\rho$) which is  -- as any norm should be -- invariant under
representation in any orthonormal basis. Let us stress that the condition ${\cal C}^{(n)} \leq 1$ guarantees
local hidden variable model for any results of binary measurements with $2^{n-1} \times 2^{n-1} \times ... \times 2^{n-2} \times ... \times 2$
settings. We have found the analytical formulas for the correlation tensor elements $\{ T_{k_{1}...k_{n}}^{2} \}$
(they are complicated and will be analysed elsewhere).
We have calculated the upper bound $||T||_{HS}^{2}$ from (\ref{bound}) and it happens that for some specific states $\rho_{n,k}$
it is smaller than one despite the fact that quantum Fisher information detects its entanglement (see main text).

{\it Sub-shot noise limit with much weaker bound entanglement.}
The following states:
\begin{equation}
\rho_{n,k,m}=\lambda' P_{n,k}^{+} + \frac{\lambda'}{2} \sum_{j=k}^{m}(Q_{n,j}^{+} + Q_{n,j}^{-})
\end{equation}

where  $\lambda'=((\sum_{j=0}^{k+m}\binom{n}{j})^{-1}$ satisfies PPT test for
any cut $j:n-j$ with $j\leq m+1$.
The corresponding Fisher information is bounded by
\begin{equation}
F_Q^{n,k,m}\geq (n-2k)^{2}\frac{k(k+1)...(k+m)}{m(n-k)(n-k-1)...(n-(k+m))}.
\label{lower-bound-app}
\end{equation}
Putting linear scaling of $k(n)=c n$ and sublinear scaling of $m(n)=n^{1-\epsilon}$
(which makes the corresponding entanglement much weaker than that
of previous states where only 1 qubit cuts obey PPT test) in the above
leads to the sub-shot noise behaviour of the Fisher information $F_Q^{n,k(n),m(n)} \sim n^{1+\epsilon}$.

\end{document}